\documentstyle[preprint,aps]{revtex}
\begin{document}
\draft
\title{\bf {Phase and periodicity of Aharonov-Bohm oscillations: effect
of channel mixing}}
\author{P. Singha Deo\cite{eml} }
\address{Dept. of Physics, University of Antwerp (UIA),\\
Universiteitsplein 1, B-2610 Antwerpen, Belgium.}
\maketitle
\begin{abstract}
We take a negative delta function impurity in one arm of a quasi one
dimensional Aharonov-Bohm ring and demonstrate abrupt phase changes
across a quasi bound state of the negative delta function potential.
We give a new mechanism for conductance oscillations with the
strength of the negative delta potential.  We also show that coupling
to evanescent modes can result in a $hc/2e$ flux periodicity at certain
energies. These observations were made in a recent ingenious
experiment by D. Mailly et al [1].
\end{abstract}
\narrowtext
\newpage

In a recent experiment [1] the conductance oscillations of an
Ahoronov-Bohm ring with a gate voltage in one of the arms is studied.
The main results of the experiment can be stated as follows. As the
gate voltage is varied the phase of Aharonov Bohm oscillations
undergo abrupt phase changes of $\pi$. Also at some gate voltages the
Aharonov-Bohm oscillations show $\phi_0/2$ periodicity, where
$\phi_0=hc/e$. The conductance is found to oscillate with the gate
voltage and this is believed to be not due to the fact that the gate
voltage changes the wave vector under the gate.
A theoretical calculation is also presented [1] that shows
such abrupt phase changes. A phase difference of $\pi$ between
successive channels added to the wavefunction of electrons as it
passes the gate in one of the arms produces such phase changes as
well as the $\phi_0/2$ periodicity in the theoretical calculation.
Such a phase difference of $\pi$ is an assumption.  If the
phase difference is made random instead of $\pi$, the agreement is
not good. But what is completely neglected in the theoretical
calculations is the fact that the gate voltage can lead to mixing of
the transverse modes. 

In this work we show that in absence of such phase randomization,
mixing of transverse modes alone can give rise to the results
observed in the experiment.  In our calculations, inorder to
demonstrate this we put a negative delta function potential in the
upper arm of the ring.  We chose the Fermi energy such that only one
mode is propagating and the others are evanescent. We consider only
one such evanescent mode but the effect of other evanescent modes
will be included in the end. The negative delta function potential in
1D has only one bound state. But in a quasi 1D wire with 2 transverse
channels, it has 2 bound states, one belonging to each channel. The
separation between these bound states is therefore same as that of
the channel quantization. The bound states that lie below the
propagating threshold are true bound states. Whereas if the bound
state of the second transverse channel is degenerate with the first
subband scattering channel, then it is a quasi bound state [2]. When
incident energy matches with this quasi bound state we get a Fano
resonance. Such resonances are characterized by a zero pole pair[3].
Across the zeros of this resonance the phase of Aharonov-Bohm
oscillations will change by $\pi$. We also give a completely new
mechanism of conductance oscillations with the strength of the delta
potential. A delta function potential cannot modify the wave vector.
Besides we show that coupling to evanescent modes can result in
$\phi_0/2$ periodicity of conductance oscillations.

A negative gate voltage in a small region narrows the thickness of
the channel over this small region. But due to box quantization quasi
bound states are formed in this narrow region [4]. Now such a quasi bound
state belonging to one subband can be degenarate with a scattering
state of another subband because although their energy is the same,
their wave vectors will depend on the zero point energy of the
respective subbands, and can be very different. And then we can observe
the same phenomenon as that observed with our negative delta
potential. As the gate voltage is changed it changes the dimension of
the narrow region, thus shifting the resonances in energy. If the
length of the gate is small then the separation between the
resonances in the begining will be same as that of channel quantization.
Besides impurities under the gate voltage can behave as a negative
delta function potential of slowly varying strength. In fact any bound
state in the system due to some reason, under the gate can cause the
effects reported here.

A schematic diagram of the Aharonov Bohm ring with a delta function
potential at site X is shown in fig. 1. Various regions and various
length parameters are also defined in the figure.  Except for the
delta function potential the potential inside the system is zero
everywhere. Distance of the delta potential from the central line
between the inner and outer radius of the ring is $y_i$. At the edges
of the system the potential is infinite which enforces hard wall
boundary conditions at the edges of the system. The widths of the
quantum wires making up the system is $w$.  Scattering properties and
bound states of a delta function potential in a quasi 1D wire has
been solved by Bagwell [5]. Since along the width of the wires the
system has perfect symmetry, different subbands will never mix,
except at the delta function potential. Some amount of mixing can be
there at the junctions of the leads but that will not cause any
additional effect because the scattering effects of junctions can be
included in the scattering effects of the delta function potential
(see eqn 2 in Ref. [1]).  So the evanescent mode in the lower arm of
the ring and those in the leads drop out from our calculations.
Since only two subbands of opposite parity are considered here, it
also drops out in the upper arm if the delta function potential is
situated at the center along the width of the upper arm, i.e.,
$y_i=0$.  So by shifting the position of the delta function potential
in the upper arm we can include or exclude the role of channel
mixing. The solutions of Schrodinger equation, in the absence of
magnetic field, in the various regions are written down below. The
choice of coordinates and origin of coordinates is easy to understand
but the conventions are the same as that in ref [6] and [7]. We have
set $\hbar=2m=1$.

\begin{equation}
\psi_I^p(x,y)=\xi_1(y)(e^{ikx}+Re^{-ikx})
\end{equation}

\begin{equation}
\psi_{II}^p(x,y)=\xi_1(y)(Ae^{ikx}+Be^{-ikx})
\end{equation}

\begin{equation}
\psi_{II}^e(x,y)=\xi_2(y)(Ce^{-qx}+De^{qx})
\end{equation}

\begin{equation}
\psi_{III}^p(x,y)=\xi_1(y)(Ee^{ikx}+Fe^{-ikx})
\end{equation}

\begin{equation}
\psi_{III}^e(x,y)=\xi_2(y)(Ge^{-qx}+He^{qx})
\end{equation}

\begin{equation}
\psi_{IV}^p(x,y)=\xi_1(y)(Je^{ikx}+Ke^{-ikx})
\end{equation}

\begin{equation}
\psi_{V}^p(x,y)=\xi_1(y)Te^{ikx}
\end{equation}

Here the superscripts `$p$' and `$e$' stands for `propagating' and
`evanescent' modes, respectively. Also here $\xi_n(y)=\sin{n\pi \over
w}(y+w/2)$, $k=\sqrt{En-E1}$ and $q=\sqrt{E2-En}$, where $w$ is the
width of the quantum wires, $En$ is the incident energy,
$E1=(\pi/w)^2$ is the propagating threshold of the first subband and
$E2=(2\pi/w)^2$ is the propagating threshold of the second subband.
We use the three way splitter of ref [8] to match boundary conditions
at the junctions $J_1$ and $J_2$, and the formalism of Bagwell to
match the wave functions across the delta function potential.
Magnetic field is treated in the way described in ref [6]. The
expression for transmission coefficient across the whole system is
too long to be produced here. However below we give an analytical
expression for the phase $\theta$ of transmission amplitude across a
delta function potential in a quantum wire with one propagating mode
and one evanescent mode in the wire.

\begin{equation}
\theta=Arctan[{-q\gamma \sin^2{\pi \over w}(y_i+w/2) \over k(2q+ \gamma
\sin^2{2\pi \over w}(y_i+w/2)}]
\end{equation}

Here $\gamma$ is the strength of the delta function potential.  It
must be noted that this expression is valid only if $y_i \ne 0$ or
else the evanescent mode gets decoupled from the propagating mode and
the problem reduces to that of a delta potential in 1D. The condition
for the bound state of the second subband for the delta function
potential is given by

\begin{equation}
(2q+ \gamma \sin^2{2\pi \over w}(y_i+w/2)=0
\end{equation}

\noindent and when the incident energy satisfy this condition
$\theta$ abruptly changes by $\pi$. Such abrupt phase changes may
have two fold consequences. It can lead to violation of Leggett's
conjecture [9] and can also lead to abrupt phase changes of
conductance oscillations. To check if it leads to violation of
Leggett's conjecture one has to check if there are any
discontinuities in Real[1/T] [9]. One can check that Real[1/T]=1 for
all energies and so there is no discontinuity. However the above
mentioned phase $\pi$ does cause an abrupt change in the phase of
conductance oscillations.  We choose $w$ as the unit of length and
all other length units are scaled by this. All energies are also
scaled by $1/w^2$. So from now on we will only give their numerical
values without mentioning the units.  If we choose $l_1=l_2=.25$,
$l_3=.5$, $\epsilon=4/9$, [10] $\gamma$=-13.968446982793 and
$y_i=w/3$ then the bound state occur at En=12 which is in between E1
and E2. In fig. 2 we show that conductance oscillations with
$\alpha=2\pi \phi / \phi_0$, where $\phi$ is the flux through the
ring, at En=12-.1 and En=12+.1 are of opposite phases. This happens
for infinitesimal $y_i$ or infinitesimal mixing of modes. The
conductance oscillations are so small in amplitude because we are
very close to a zero in the transmission coefficient of the upper
arm. For all $En$ on the higher (lower) side of this value the
conductance oscillations are in phase.  Hence when every subband has
a quasi bound state separated by the typical channel quantization
value, oscillatory behavior in conductance with $\gamma$ is a
straightforward conclusion.

We also find another striking feature of this coupling to evanescent
modes. In absence of any mode coupling, i.e., when the system reduces
to a 1D system, the conductance oscillations have a $\phi_0$
periodicity. Most of the time they exhibit one maxima(or minima) at
$\alpha$=0 and one minima (or maxima) at $\alpha=\pi$. But then at
some energies additional minimas can occur due to the resonances of
the ring [11]. But generally these additional minimas are very
different in strength (or amplitude) compared to the others. These
diversities are typical features of ballistic systems that are
extremely sensitive to boundary conditions. Coupling to evanescent
modes smooths out these diversities by reducing sensitivity to
boundary conditions. We find a very general feature of the coupling
to evanescent modes is that it makes the oscillations of comparable
strengths. And this in turn gives the conductance fluctuations the
appearance of $\phi_0/2$ periodicity. A typical example of this is
shown in fig 3.  The dotted curve is plotted for $l_1=l_2=.25$,
$l_3$=.5, $y_i=0$, $En=26.5$, $\gamma=-13.968446982793$ and
$\epsilon=4/9$.  Whereas the solid curve is plotted for the same
parameters except that $y_i=w/2.5$. So for the dotted curve there is
no coupling to evanescent modes whereas for the solid curve there is.
The solid curve has almost exact $\phi_0/2$ periodicity. If we move
away from this incident energy then again the periodicity slowly
changes.  Given a position of $y_i$ one can play with the incident
energy to find a value where the same happens. Or else one can keep
incident energy fixed and vary $\gamma$ to get it. They are
equivalent.  Another set of parameter values for which this happens
are mentioned below. If $l_1=l_2=.25$, $l_3$=.5,
$\gamma$=-13.968446982793, $\epsilon$=4/9 and $y_i=w/3.5$ then this
happens around an energy of $En=24.5$.

The effect of other evanescent modes in the system will be to
renormalize the strength of coupling between the two subbands
considered here, making channel mixing stronger, apart from giving
rise to more quasi boundstates that are separated by the typical
channel quantization value. The scheme of renormalization is given in
ref 5.

Hence most of the experimental observations are obtained in a simple
model by considering existence of quasi bound states and coupling to
evanescent modes. It is not necessary to assume a phase difference of
$\pi$ between successive channels.  It also shows a new mechanism for
oscillatory behaviour with gate voltage.  The study also justifies
that the phase changes in conductance oscillations obtained by Yacoby
et al [12] are similar and arise due to elastic scattering and Fano
resonances associated with the special geometry in that case [13].

I thank Prof. F. Peeters for useful discussions.  I acknowledge with
gratitude the discussions I had with Prof. Mailly when I was visiting
his laboratory in Nov. 1996. I also thank him for communicating to me
the results of the experiment and for giving me a rough write up of
the work. This work is supported by a scholarship from the University of
Antwerp and support from the FWO-V grant No: G.0232.96.

\vfill
\eject

\vfill
\eject
\centerline{FIGURE CAPTIONS}

Fig. 1  A quasi one dimensional Aharonov-Bohm ring with a delta
function potential at site marked X in the upper arm of the ring. A
magnetic flux penetrates the ring.

Fig. 2  Plot of transmission coefficient versus $\alpha=2 \pi \phi
/\phi_0$ for parameter values described in the text, for two values
of incident energies on opposite sides of a quasi bound state.

Fig. 3  Plot of transmission coefficient versus $\alpha=2 \pi \phi /
\phi_0$  with no channel coupling (dashed line) and with channel
coupling (solid line).


\begin{thebibliography}{99}
\bibitem[*]{eml} deo@uia.ua.ac.be
\bibitem{mai} D. Mailly et al (unpublished).
\bibitem{ban} Hence these bound states cannot capture electrons like
a quantum dot.
\bibitem{tek} E. Tekman and P. F. Bagwell, Phys. Rev. B {\bf 48},
2553(1993).
\bibitem{sto} A. Szaffer and A. D. Stone, Phys. Rev. Lett. {\bf 62},
300(1989)
\bibitem{bag} P. F. Bagwell, Phys. Rev. B {\bf 41}, 10354(1990).
\bibitem{deo2} A. M. Jayannavar and P. Singha Deo, Mod. Phys. Lett. B
{\bf 8}, 301(1994).
\bibitem{deo3} P. Singha Deo, B. C. Gupta and A. M. Jayannavar,
(manuscript under preparation).
\bibitem{but} M. B$\ddot u$ttiker, Y. Imry and M. Azbel, Phys. Rev A
{\bf 30}, 1982(1984).
\bibitem{deo4} P. Singha Deo, Phys. Rev. B {\bf 53}, 15447(1996).
\bibitem{deo5} For this value of $\epsilon$ the three way splitter
becomes exact and it corresponds to free junctions. See P. Singha Deo
and A. M. Jayannavar, Mod. Phys. Lett. {\bf 7}, 1045(1993).
\bibitem{pas} J. D'Amato, H. M. Pastawski and J. F. Weiss, Phys. Rev.
B {\bf 39}, 3554(1989).
\bibitem{yac} A. Yacoby, M. Heiblum, D. Mahalu and H. Shtrikman,
Phys. Rev. Lett. {\bf 74}, 4047(1995).
\bibitem{deo1} P. Singha Deo and A. M. Jayannavar, Mod. Phys. Lett.
B {\bf 10}, 787(1996). 
\end{thebibliography}
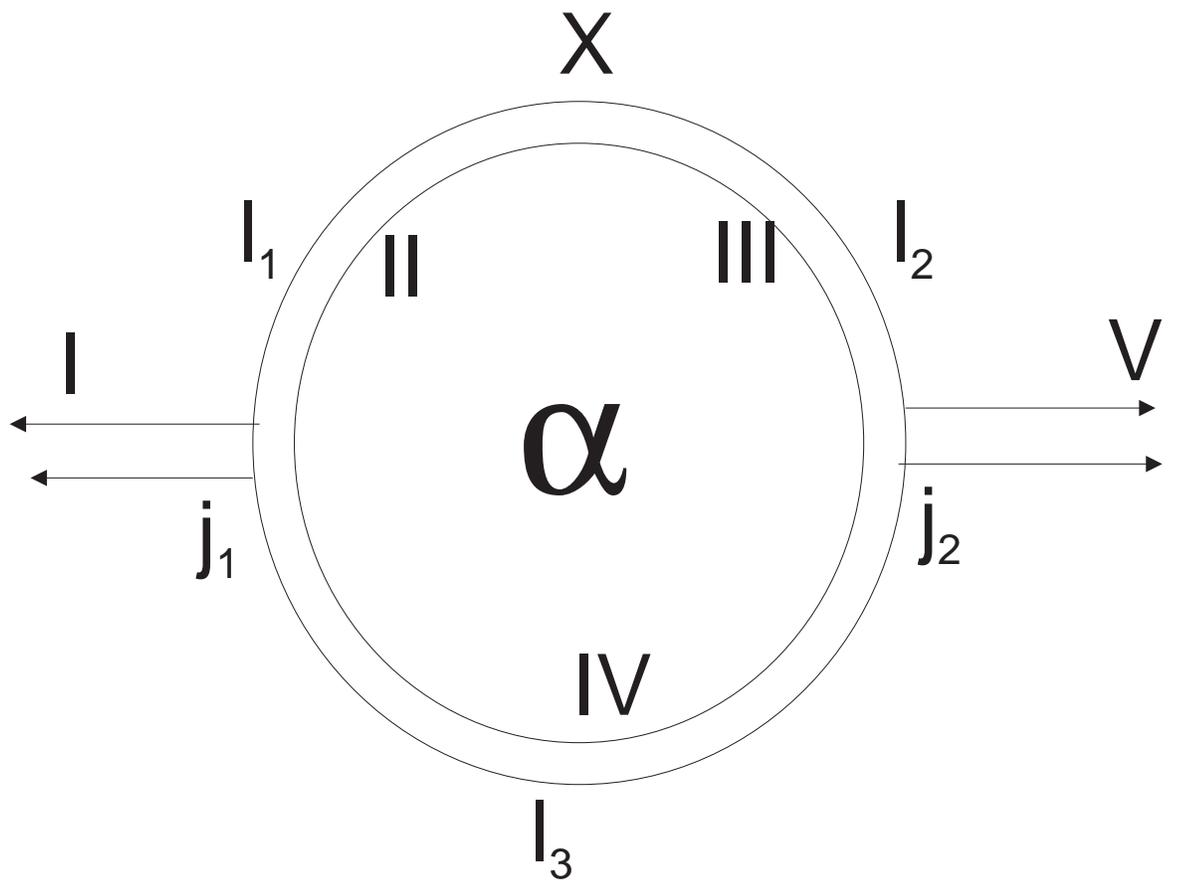
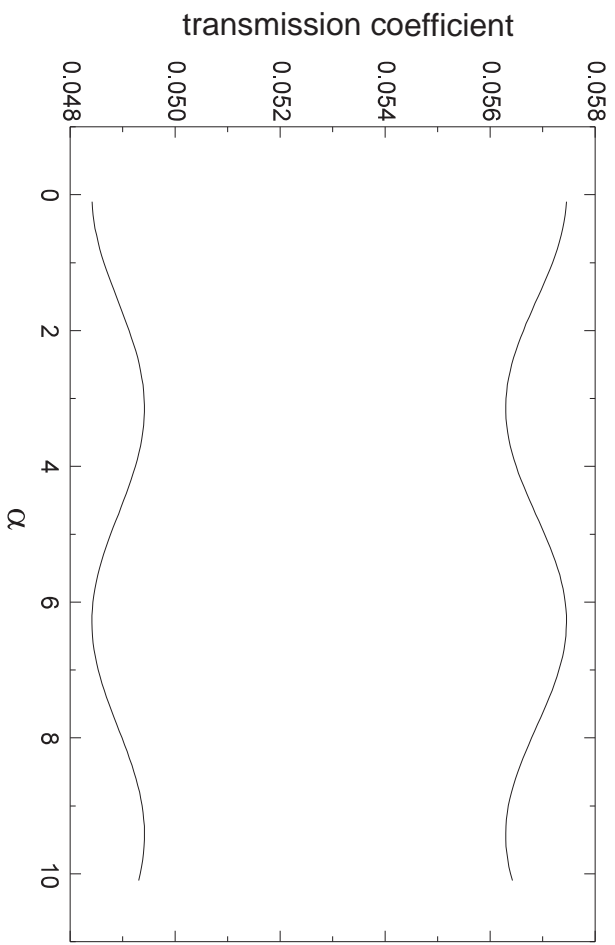
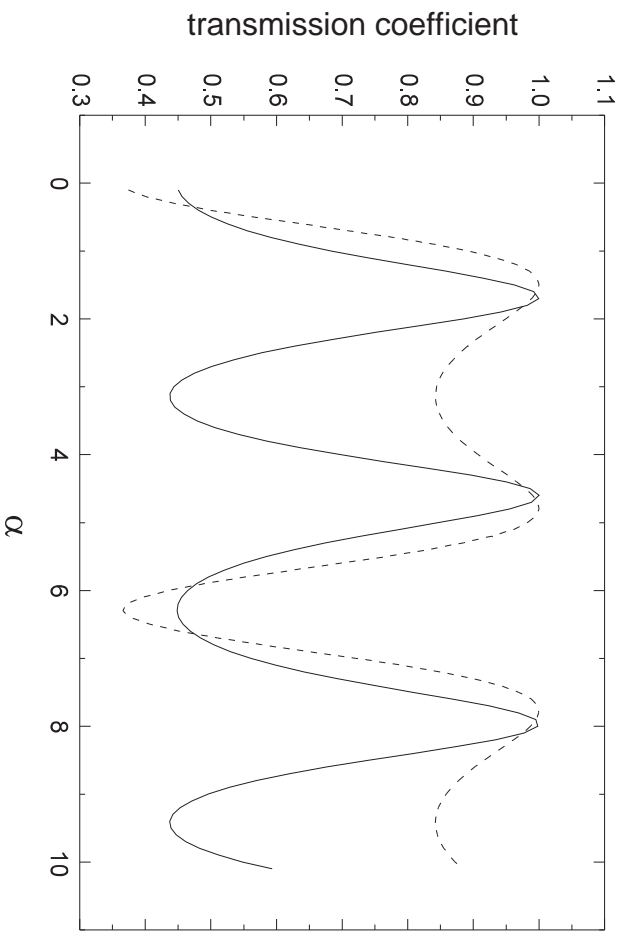
\end{document}